\documentclass[aps,twocolumn,superscriptaddress,altaffilletter,lengthcheck,
tightenlines,showpacs,showkeys]{revtex4}
\usepackage{color}

\newcommand{\al}{\alpha}

\newcommand{\bb}{\beta}

\newcommand{\D}{\Delta}
\newcommand{\ben}{\begin{eqnarray}}

\newcommand{\een}{\end{eqnarray}}

\newcommand{\be}{\begin{equation}}

\newcommand{\ee}{\end{equation}}

\newcommand{\n}{\label}

\newcommand{\no}{\noindent}

\newcommand{\ro}{\rho}

\newcommand{\om}{\omega}

\newcommand{\Om}{\Omega}

\usepackage{graphicx}
\usepackage[dvipdf]{epsfig}

\usepackage{color}

\begin{document}

\title{Modified holographic Ricci dark energy  coupled to interacting  dark matter\\
 and a non interacting  baryonic component}

\author{Luis P. Chimento}\email{chimento@df.uba.ar}
\affiliation{Departamento de F\'{\i}sica, Facultad de Ciencias Exactas y Naturales,  Universidad de Buenos Aires and IFIBA, CONICET, Ciudad Universitaria, Pabell\'on I, 1428, Buenos Aires  , Argentina.}
\author{M\'onica Forte}
\email{monicaforte@fibertel.com.ar}
\affiliation{Departamento de F\'isica, Facultad de ciencias Exactas y Naturales,
Universidad de Buenos Aires, Ciudad Universitaria, Pabell\'on I, 1428, Buenos Aires  , Argentina.}
\author{Mart\'{\i}n G. Richarte}\email{martin@df.uba.ar}
\affiliation{Departamento de F\'{\i}sica, Facultad de Ciencias Exactas y Naturales,  Universidad de Buenos Aires and IFIBA, CONICET, Ciudad Universitaria, Pabell\'on I, 1428, Buenos Aires  , Argentina.}

\begin{abstract}
We examine a Friedmann-Robertson-Walker universe  filled with interacting dark matter,  modified holographic Ricci  dark energy (MHRDE), and a decoupled  baryonic  component.The estimations of the cosmic parameters with Hubble data lead to an age of the universe of $13.17 {~\rm Gyr}$ and show that the MHRDE is free from the cosmic-age problem at low redshift ($0 \leq z \leq 2$) in contrast to holographic Ricci dark energy (HRDE) case. We  constrain the parameters with the Union2 data set and contrast with  the Hubble data. We also study  the behavior of dark energy at early times by taking into account the severe bounds found at recombination era and/or  at big bang nucleosynthesis. The inclusion of a non interacting baryonic matter  forces that the amount of dark energy at  $z_{t} \sim {\cal O}(1)$ changes abruptly implying that $\Omega_x (z\simeq 1100) =0.03$, so  the bounds reported by the forecast of  Planck  and CMBPol experiments are more favored for the  MHRDE model than  in the case of HRDE cutoff. For the former model, we also obtain that at high redshift  the fraction of dark energy varies from $ 0.006$ to $ 0.002$, then the amount of $\Omega_x $ at the big bang nucleosynthesis era  does not disturb the observed Helium abundance in the universe provided that the bound $\Omega_{x}(z\simeq 10^{10})<0.21$ is hold.

\end{abstract}

\vskip 1cm



\date{\today}

\bibliographystyle{plain}

\maketitle

\vskip 8cm
\section{Introduction}

The holographic principle states that  the maximum  number of degrees of freedom  in a volume should be proportional to the surface area \cite{holo1}, \cite{holo2}, \cite{holo3}, \cite{holo4}.  Using the effective  quantum field theory  it was shown that  the zero-point  energy  of a system  with size $L$ should no exceed  the mass  of a black hole with the same size, thus $L^{3}\ro_{\Lambda}\leq LM^{2}_{P}$, where $\ro_{\Lambda}$ corresponds to the quantum zero-point energy density \cite{holo5} and $M^{-2}_{P}= 8\pi G$. The latter relation establishes a link between  the ultraviolet cutoff, define through $\ro_{\Lambda}$, and the infrared cutoff  which is encoded by the scale $L$. Applying this novel principle within the cosmological context implies that the dark energy density of the universe $\ro_{x}$ takes the same form of the vacuum energy, $\ro_{\Lambda}=\ro_{x}$. Using the largest $L$ as the one saturating the above inequality, it turns out to be the holographic dark energy is given by $\rho_{x}=3c^{2}M^{2}_{~P}L^{-2}$, where $c$ is a numerical factor. The IR cutoff  has been  taken  as the large scale of the universe, Hubble horizon \cite{hde0,hde1}, particle horizon, event horizon \cite{hde1} or generalized IR cutoff \cite{hde2}, \cite{ricci1}, \cite{ricci2}, \cite{ricci3} \cite{hv}, \cite{hdiego}, \cite{hmi1}, \cite{hmi2} amongst many others. One of the main reasons in working within the framework of dynamical dark energy such as a HDE relies on the need of explaining the current accelerated phase of the Universe; this fact has been confirmed by a plethora of observational  tests such as high redshift Hubble diagram of type Ia supernovae as standard candles \cite{c1a,c1b,c2a,c2b} and  accurate measurements of  cosmic microwave background (CMB) anisotropies\cite{c4a,c4b,c4c}. In order to play the role of a dynamical dark energy model, the infrared cutoff will be  considered  as a function of the cosmic time so the holographic dark energy will evolve dynamically. Here, we will focus our attention on an extended version of the well known Ricci scalar cutoff \cite{ricci2}. An important feature of this model refers to the fine-tuning problem, that is, due to the dark energy density is based on space time scalar curvature, without involving a Planck or high physical energy scale, the fine-tuning problem is avoided and the coincidence problem is also discarded within this context \cite{ricci1}. Several works have been devoted to obtain cosmological constraints  on the holographic Ricci dark energy model \cite{ricciobse1}, \cite{ricciobse2}, \cite{hmi1}, \cite{hmi2}, \cite{RMObserva} or generalized versions of the latter one \cite{ricciobse4}. Nevertheless dark energy is not the only mysterious element in the Universe, the necessity of a dark matter component comes from astrophysical evidences of colliding galaxies, gravitational lensing of mass distribution or power spectrum  of clustered matter \cite{DMobserva}, \cite{dme}. Moreover, the astrophysical observations from the galactic to the cosmological scales indicate that dark matter is a substantial component to the universe's total matter density, being responsible for the structure formation in the Universe \cite{DMobserva}.  Presents attempts to understand the physics behind the dark sector composed of dark matter and dark energy have indicated that  
there is an unavoidable degeneracy between dark matter and dark energy within Einstein's gravity, namely, there could be a hidden non-gravitational coupling between them without violating current observational constraints and thus it would be interesting to develop ways of testing exchange of energy in the dark sector. More precisely,  if dark energy interacts with dark matter, there is a change in the background evolution of the Universe that allows us to constrain a phenomenological type of interaction.  Therefore, an holographic scenario becomes a physically viable model when one takes into account a possible interaction between the dark matter and dark energy. It entails that the dark matter feels the presence of  the dark energy through the  gravitational expansion of the universe plus the  exchange of energy between them.  In fact, we will follow a phenomenological approach by  studying the properties hidden in a particular kind of  interaction  and then one confronts  the theoretical model with the available observational data. Recently several known linear and nonlinear interactions in the dark sector have been generalized\cite{jefe1} . It was introduced  an effective one-fluid description  of the dark components and shown  that interacting  and unified models are related to each other. It should be stressed that interacting dark energy scenarios have been studied by many authors \cite{jefe1}, \cite{InteraDE}.

It is well known that some new physics may be showing up at high redshift taking into account strict limits coming from big bang nucleosynthesis (BBN) data. More precisely, the $^{4}He$ abundance has often been used as a sensitive probe of new physics. This is essentially due  to the fact that  nearly all available neutrons at the time of BBN, in a scale of $1$Mev of temperature, at the time of BBN end up in  $^{4}He$ and the neutron-to-proton ratio is very sensitive to the competition between the weak interaction rate and the expansion rate. For example, a bound on the number of relativistic degrees of freedom (d.o.f), $g_{*}$, at the time of BBN commonly known as the limit on neutrino flavors, $N_{\nu}$, is derived through its effect on the expansion rate, $H \propto \sqrt{g_{*}} T^{2}$ where $g_{*}=2+7/2 + 7N_{\nu}/4$ counts the relativistic dof in photons, $e^{\pm}$ pairs, and  $N_{\nu}$ neutrino species; having assumed a Universe dominated by radiation so that $\ro \propto g_{*} T^{4} $\cite{Cyburt}. Moreover, the presence of vacuum energy during BBN is well motivated both by considerations of dark energy as well as inflation, giving as stringent bound $\Omega_{x}(1 \mbox{Mev})<0.21$ \cite{Cyburt}. Besides, the physics at  recombination era gives also some constraints in the amount of dark energy at such primordial era which also has to be consistent with the severe bounds provided by BBN data mentioned above. In particular,  stringent signal could arise from the early dark energy (EDE) models, that is, uncovering the nature of dark energy as well as their properties  to high redshift along with their effects imprinted on the universe could provide invaluable guide to the physics behind the recent speed up of the universe \cite{EDE1}. Therefore,  any  serious dark 
energy model used for constraining the present-day value of $\Omega_{x}(z=0)$  also has to be consistent 
with the early bounds on the fraction of dark energy at primordial eras such as recombination or BBN.
Not too long ago it was examined the current and future data  for constraining the amount  of EDE, the cosmological data
analyzed has led to an upper bound of  $\Omega_x(z \simeq 1100)<0.043$ with $95\%$ confidence level (C.L.) in case of relativistic EDE while for a 
quintessence type of EDE has given $\Omega_x(z \simeq 1100)<0.024$ although   the EDE component is not preferred, it is also not 
excluded from the current data \cite{EDE1}. Another appealing forecast for  the bounds of the EDE taking into account Planck  and CMBPol 
experiments can be found in \cite{EDE2}. More precisely, assuming a $\Omega_{x}(a \simeq 10^{-3}) \simeq 0.03$ among other priors, it 
was checked the stability of these values, interesting enough  was the  $1\sigma$ error coming from Planck experiment 
giving  as result $\sigma^{Planck}_{x} \simeq 0.004$ whereas the CMBPol  improved this bound by a factor 4 $\sigma^{CMBPol}_{x}\simeq 10^{-3}$ \cite{EDE2}. Besides, some new limits on EDE from the CMB using the data from the WMAP satellite on large angular 
scale and the South Pole Telescope (SPT) on small angular scale were obtained in \cite{EDE3}. Considering the CMB data alone 
it got a bound of $\Omega_x(z \simeq 1100)<0.0018$ very similar to the one reported in \cite{EDE2}. In addition,  the constraints on the variation in the fine structure constant \cite{EDE4} in the presence of EDE gave an upper bound of $\Omega_x(z \simeq 1100)<0.06$ at $95\%$ C.L. which is weaker than the bounds reported in \cite{EDE1}-\cite{EDE2}. 

The present article is outlined as follows. We investigate a universe composed of interacting dark matter, modified holographic Ricci dark energy(MHRDE), where the exchange of energy in the dark side is proportional to derivative of total dark sector energy density, and a decoupled component that could behave as baryonic matter at early times. We use the Hubble data and the Union2 compilation of SNe Ia  for constraining the cosmological parameters, thus we compare phenomenological aspects of MHRDE and HRDE holographic dark energy models. We also make a kinematic analysis  for studying the behavior of decelerating parameter, equations of state and the ratio dark matter to dark energy. Using the best fit values, we estimate the age of the universe and explore the cosmic age-problem.  As a complementary tool for getting more accurate constraints on both models, we examine the behavior of dark energy at early times. 

\section{The model }
We assume a flat FRW universe  filled with three different components, an interacting dark sector composed of a  nearly pressureless dark matter, MHRDE, and a decoupled  baryonic contribution  with energy densities $\ro_c$, $\ro_x$, and $\ro_{b}$,  respectively.  We adopt as  equations of state  $\omega_c=p_c /\ro_c$ for dark matter, $\omega_x=p_x/ \ro_x$ for dark energy, and $\omega_b=p_b/ \ro_b$ for radiation, thus the Einstein equations  read
\be
\n{01}
3H^{2}= \ro_c + \ro_x+\ro_b,
\ee
\be
\n{02a}
\dot\ro_c + \dot\ro_x +3H[(\omega_c+1)\ro_c + (\omega_x +1 )\ro_x]=0,
\ee
\be
\n{02b}
\dot\ro_b + 3H(\omega_b+1)\ro_b =0,
\ee
where $a$  is the scale factor, $H = \dot a/a$ stands for the Hubble expansion rate. Here, we will use the holographic principle within the cosmological context  by associating the infrared cutoff $L$ with the dark energy density, thus we take $L^{-2}$ in the form of a linear combination of $\dot{H}$ and $H^{2}$ \cite{ricci2}:
\be
\n{03}
\ro_x=\frac{2}{\al -\bb}\left(\dot H + \frac{3\al}{2} H^2\right),
\ee
being  $\al$ and $\bb$ two free constants. In particular, we obtain $\ro_x\propto R$ for  $\al=4/3$ \cite{hdiego}, where $R=6(\dot{H}+2H^{2})$ is the Ricci scalar curvature for a spatially flat FRW space-time.

The use of the variable $\eta = \ln(a/a_0)^{3}$, where $a_0$ is set as the value of the scale factor at present, allows us to rewrite Eqs. (\ref{02a})-(\ref{03}) as
\be
\n{04}
\ro=\ro_c+\ro_x,
\ee
\be
\n{05}
\ro'=-(\omega_c+1)\ro_c -(\omega_x+1)\ro_x,
\ee
\be
\n{06}
\ro'=-\al\ro_c -\bb\ro_x,
\ee
\be
\n{06b}
\ro'_b=-(\omega_b+1)\ro_b,
\ee
where the prime stands for derivatives with respect to the new variable $' \equiv d/d\eta$ and the condition $0<\beta<\al$  is imposed to avoid a phantom scenario. From Eq. (\ref{06b}) is clear that the radiation component is decoupled from interacting dark sector, so the exchange of energy  only takes place between the dark matter and dark energy, thus $\ro_{b}=\ro_{b0}a^{-3(\omega_b+1)}$ and its density parameter is $\Omega_{b}=\ro_{b0}a^{-3(\omega_b+1)}/(3H^{2})$.

MHRDE (\ref{03}), with a term proportional to $\dot H$ leads to Eq. (\ref{06}), which looks like  a ``conservation equation" for the two dark components with constant coefficients. We will refer  to the Eq. (\ref{06}) as the modified conservation equation (MCE). Comparing the whole conservation equation (WCE) (\ref{05}) and the MCE (\ref{06}),  namely $(\om_c+1)\ro_c + (\om_x+1)\ro_x=\al\ro_c + \bb\ro_x$, we obtain the compatibility relation  
\be
\n{07}
\om_x=(\al-\om_c-1)r+\beta-1,
\ee
between the equation of state of both components and its ratio $r=\ro_c/\ro_x$. In what follows, we will use the MCE (\ref{06}) with constant coefficients $\al$ and $\beta$ instead of the WCE (\ref{05}) with  non-constant coefficients.  In some sense, the WCE (\ref{05}) and the MCE (\ref{06}) give rise to  different representations of the mixture of two interacting dark fluids and clearly these descriptions are related between them by the compatibility relation (\ref{07}). Therefore,  the MHRDE conveniently links a model of two interacting fluids having {\bf variable equations of state} with a model of two interacting fluids with ``{\bf constant equations of state}". 

Using Eqs. (\ref{04}-\ref{05}) the total pressure is $p = p_c + p_x$  and the effective equation of state of the dark sector (EOS), $\omega=p/\rho$ can be rewritten as 
\be
\n{12}
p= -\ro -\ro', \,\,\,\,\, \omega =\frac{\omega_{c}r+\omega_{x}}{1+r},
\ee

At this point, we introduce an interaction $3HQ$ between the dark components by splitting the MCE (\ref{06}) in the following way
\be
\n{08}
\rho'_{c}+\al \rho_{c}=-Q, \,\,\,\,\,\,\,\, \rho'_{x}+\beta\rho_{x}=Q.
\ee

Now, we assume a pressureless dark matter ($\om_c=0$), hence the equation of state of dark energy (\ref{07}) becomes linear in $r$
\be
\n{gac1}
\om_x=(\al -1)r+\beta-1. 
\ee

The next step is to introduce a phenomenological interaction between the dark components in order to extract some physics information about the behavior of them. We are going to study an interacting scenario where the exchange of energy between dark matter and dark energy  is  proportional to  $\ro'$. We will employ the method developed by one of the authors \cite{jefe1} based on the source equation for obtaining the total energy density of the dark sector once the interaction is given, then we will be able to reconstruct the partial energy densities. The new kind of interaction  was introduced in \cite{jefe1} and reads as
\begin{equation} 
\label{Q}
Q= -\frac{(\om_s +1 - \al)(\om_s +1 -\bb)}{(\om_s+1) \Delta}\,\ro',\label{ql}
\end{equation}
where $\om_{s}$ is a constant that varies between $\al-1$ and $\beta-1$, $\D = \al -\bb$, and $Q<0$. Taking into account that the partial energy densities $\ro_{c}$ and $\rho_{x}$ appears as a linear combination in the conservation equation(\ref{06}), the interaction (\ref{Q}) can be expressed as  a linear combination of $\ro'_{c}$ and $\rho'_{x}$ also.
In what follows we will employ the method of the ``source equation'' developed in \cite{jefe1}  for obtaining the total energy density  as
\begin{equation}
\ro=b_1a^{-3\frac{\al\beta}{(\om_s+1)}}+b_2a^{-3(\om_s+1)},\label{densf}
\end{equation}
where $b_1$, $b_2$ are integration constants. At early times the effective energy density of the dark sector takes the form  $\ro \simeq  b_1 a^{-3\al\beta/(\om_s+1)}$, in order to have a term $a^{-3}$ in the dark sector  we need to take  $\al\beta=\om_s+1$. In latter case,  the dark matter and dark density parameters  take the forms 
\be
\n{27}
\Omega_c=\frac{(1-\beta)b_1+(\om_s-\beta+1)b_2a^{-3\om_s}}{(b_{1}+ b_{2}a^{-3\om_s})\Delta}, 
\ee
\be
\n{28}
\Omega_x=\frac{(\al-1)b_{1}+(\al-\om_s-1)b_{2}a^{-3\om_s}}{(b_{1}+ b_{2}a^{-3\om_s})\Delta}.
\ee
Dark matter and dark energy densities  behave as $a^{-3}$ with a constant ratio $r_e\simeq  (1-\beta)/(\al-1)$ at early times. However, at late times the dark components behave as $a^{-3\al\beta}$ so the parameter densities (\ref{27}) and (\ref{28})  give $\Omega_c\simeq [\omega_{s}-\beta+1]/[\al-\beta]$ and $\Omega_x\simeq[ \al-1-\omega_{s}]/[\al-\beta]$, hence $r_l\simeq \beta(\al-1)/\al(1-\beta)$ and $\om_x\simeq  [\beta(\al-1)^2-\al(\beta-1)^2]/\al(1-\beta)$. 

%
\section{Cosmological Constraints}
In what follows, we will place some constraints on the model, mentioned in the last section, using the observational Hubble $H(z)$ data and the constraints imposed by the Union 2 compilation of SNe Ia. The function $H(z)$ plays a crucial role to understand the properties of the dark energy since its value is directly obtained from  astrophysical observations. More precisely, the differential age data of astrophysical objects that have evolved passivelly during the history of the universe (e.g. red galaxies) allows to test theoretical cosmological models through the predicted Hubble function $H(z) = -(1 + z)^{-1}dz/dt$ expressed in terms of the redshift $z$. Hence, we obtain the function $H(z)$ by direct determination of  $dz/dt$ \cite{obs3}. This can be achieved by identifying some ``clock'' galaxies that exhibit a uniform distribution of star population \cite{obs3}. The 12 observational  $H(z)$ data is listed in \cite{obs4}. There, $H_{obs}(z_i)$ and $H_{obs}(z_k)$ are uncorrelated because they are obtained from the observations of galaxies at different redshifts, where $z$ varies over the interval $[0, 1.75]$. 
The  statistical analysis is based on the $\chi^{2}$--function of the Hubble data which is constructed as (e.g.\cite{Press})
\be
\n{c1}
\chi^2(\theta) =\sum_{k=1}^{12}\frac{[H(\theta,z_k) - H_{obs}(z_k)]^2}{\sigma(z_k)^2},
\ee
where the $\theta$ symbol refers to the set of cosmological parameters, $H_{obs}(z_k)$ is the observational $H(z)$ data at the redshift $z_k$, $\sigma(z_k)$ is the corresponding $1\sigma$ uncertainty, and the summation is over the 12 observational  $H(z)$ data listed in \cite{obs4}, \cite{obs4b}. From this quantity, the probability distribution function (PDF) is constructed as ${\cal P}={\cal W} e^{-\chi^{2}/2}$ where ${\cal W}$ is a normalization factor. 

The $\chi^2$--function will be  minimized for obtaining the best-fit values of  the random variables $\theta_{c}$ that correspond to a maximum of ${\cal P}$. The  best fit parameters $\theta_{c}$ are those values where $\chi^2_{min}(\theta_{c})$ leads to a local minimum of the $\chi^2(\theta)$--distribution. If $\chi^2_{d.o.f}=\chi^2_{min}(\theta_{c})/(N -n) \leq 1$ the fit is good and the data are consistent with the considered model $H(z;\theta)$\cite{Press}, where $N$ indicates the number of observational data whereas $n$ counts the number of parameters so $\chi^2$--function  has $N-n$ degrees of freedom. 
To better understand the cosmological constraints coming from $\chi^2$--statistical method employed here, we are going to place  constraints over  all paremeters, taken in pairs,  while the others are taken as priors, namely, we consider one pair of unknown parameters and obtain their mean value, then we choose another pair of parameters and repeat the process until all the set of $\theta$--parameters have been properly estimated. From the latter analysis,  we are going to obtain the $68.3\%$ and $95.4\%$ confidence level (C.L.) of probabibility that in the case of two independent parameters corresponds to the random data sets which satisfy the inequality $\Delta\chi^{2}=\chi^2(\theta)-\chi^{2}_{min}(\theta_{c})\leq 2.30$ and $\Delta\chi^{2}\leq 6.17$ respectively, these contours are usually closed ellipses. Here, $N = 12$, $n=2$, the string of parameters is $\theta= (H_0, \al, \bb,\Om_{c0}, \Om_{x0},\Om_{b0})$, and the theoretical Hubble function in terms of redshift $z$  is given by
\be
\n{22}
H(z) = H_0 [{\cal A} (1+z)^3 + {\cal B}(1+z)^{3(\om_s + 1)}+\Om_{b0}(1+z)^{3\al}]^{1/2}
\ee
where the constants ${\cal A}$ and ${\cal B}$ are written in term of  parameters as
\begin{eqnarray}
\label{23a}
{\cal A}&=&\frac{(\om_s + 1 - \al)\Om_{c0} + (\om_s + 1 - \bb)\Om_{x0}}{\om_s}
\\
\label{23b}
{\cal B}&=&\frac{(\al - 1)\Om_{c0} + (\bb - 1)\Om_{x0}}{\om_s}
\end{eqnarray}
having used the standard definition of the density parameters $\Om_{i0}=\rho_{i0}/3H_0^2$ with $i=\{x,c, b\}$ and the flatness condition $\Om_{c0}+\Om_{x0}+\Om_{b0}=1$ is hold. 
\begin{widetext}
\no Applying $\chi^{2}$-method to the theoretical Hubble (\ref{22}) gives us the confidence level associated with  the $1\sigma$ and $2\sigma$ probability  for all the possible pairs (see Fig. \ref{Fig: Figura9}), the best-fit value for each pair is represented by a dot and the dashed zones exclude places (in the parameter space ) where the conditions $\Om_{i0} \geq 0$  fail to be guaranteed,  the range of the parameters $\al$ and $\beta$ lead to a phantom scenario or  the parameter densities $\Omega_{i0}$ take values that are not consistent with the literature (see Fig.\ref{Fig: Figura9}).
\begin{figure}
\begin{center}
\includegraphics[height=18cm,width=12.5cm]{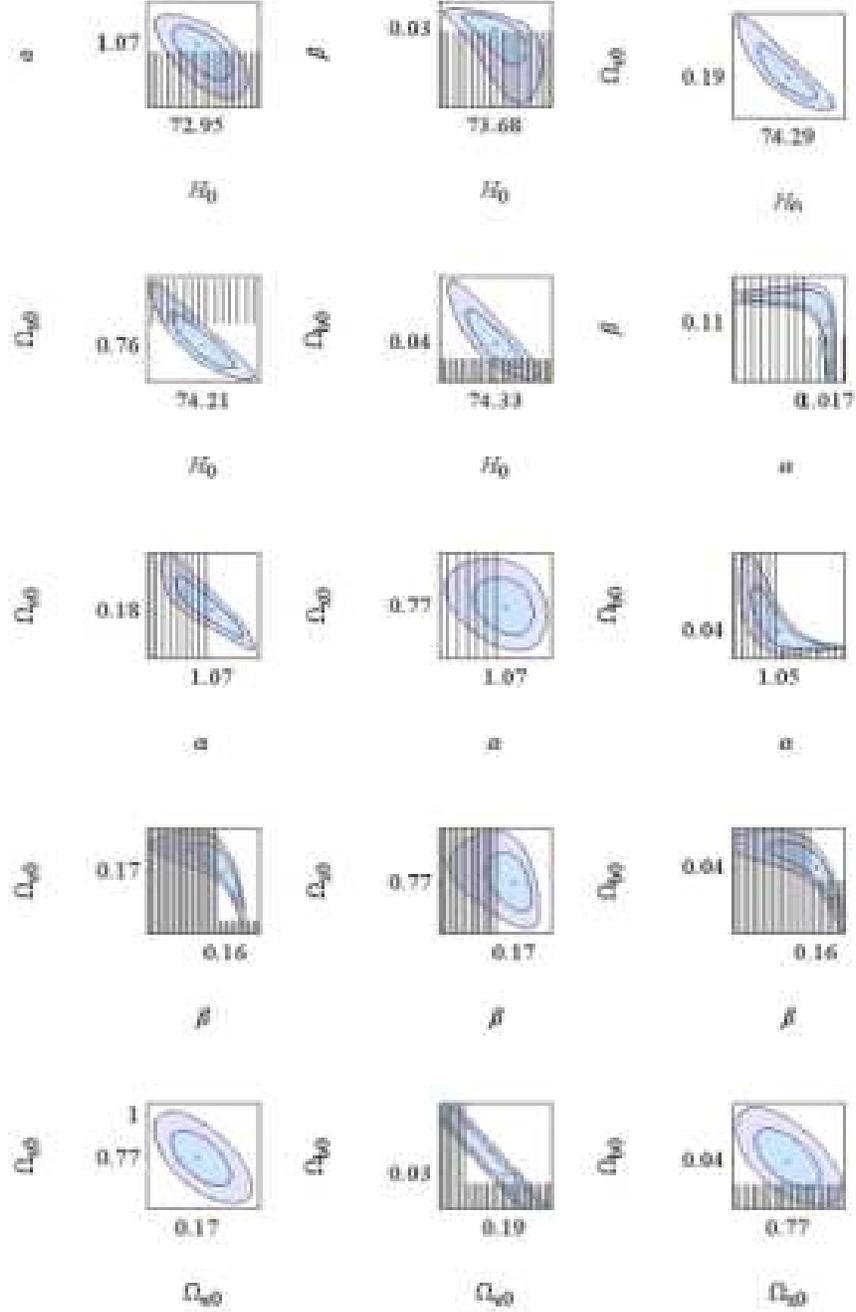}
\caption{\scriptsize{The $68.3\%$ and $95.4\%$ confidence level contours for all pairs of  $\theta= (H_0, \al, \bb,\Om_{c}, \Om_{x}, \Om_{b})$ is depicted. In each figure, the dot indicates the best fit observational value obtained with the $H(z)$ function, and the dashed bands denote the  forbidden regions for the parameters.}} 
\label{Fig: Figura9}
\end{center}
\end{figure} 
\begin{table}
\scalebox{0.6}{
\begin{tabular}{|c|c|c|c|c|c|c|c|c|c|c|c|c|c|c|c|}
\hline
\multicolumn{16}{|c|}{}\\
\multicolumn{16}{|c|}{\Large{Best fit values}}\\
\multicolumn{16}{|c|}{}\\
\hline
 & & & & & & & & & & & & & & & \\
($\theta_1$, $\theta_2$)&$(H_0,\al)$&$(H_0,\bb)$&$(H_0,\Om_{c0})$&$(H_0,\Om_{x0})$&$(H_0,\Om_{b0})$&$(\al,\bb)$&$(\al,\Om_{c0})$&$(\al,\Om_{x0})$&$(\al,\Om_{b0})$&$(\bb,\Om_{c0})$&$(\bb,\Om_{x0})$&$(\bb,\Om_{b0})$&$(\Om_{c0},\Om_{x0})$&$(\Om_{c0},\Om_{b0})$ &$(\Om_{x0},\Om_{b0})$\\
\hline
& & & & & & & & & & & & & & &  \\
($\theta_{1c}$, $\theta_{2c}$)&$(72.95,1.07)$& $(73.68,0.03)$ &$(74.29,0.19)$&$(74.21,0.76)$&$(74.33,0.04)$ & $(1.017,0.11)$&$(1.07,0.18)$& $(1.07,0.77)$& $(1.05,0.04)$&$(0.16,0.17)$&$(0.17,0.77)$&$(0.16,0.04)$&$(0.17,0.77)$&$(0.19,0.03)$&$(0.77,0.04)$ \\
\hline
$\chi_{d.o.f}^2$& 0.779& 0.769& 0.767& 0.767& 0.767 & 0.765& 0.763& 0.763 & 0.762& 0.762& 0.762& 0.762&  0.762& 0.762& 0.762  \\
\hline
\end{tabular}}
\caption{The best-fit values for each pair of parameters and the corresponding values of $\chi_{d.o.f}^2$  per degree of freedom are indicated. The average values obtained from these partial adjustments are $H_0=73.89$, $\alpha=1.05$, $\beta=0.12$, $\Omega_{c0}=0.18$, $\Omega_{x0}=0.768$ and $\Omega_{b0}=0.038$}
\label{Tabla3}
\end{table}
The best-fit values for each pair of parameters and the corresponding values of $\chi_{d.o.f}^2$  per degree of freedom are gathered together in Table \ref{Tabla3}. Up to this point,  talking in broad terms, we found that the values of $\theta$ show a small variation around their mean values: $H_0=73.89^{+0.44}_{-0.94}{\rm km~s^{-1}\,Mpc^{-1}}$, $\alpha=1.055^{+0.015}_{-0.038}$, $\beta=0.126^{+0.044}_{-0.096}$, $\Omega_{c0}=0.18^{+0.01}_{-0.01}$, $\Omega_{x0}=0.768^{+0.002}_{-0.008}$ and $\Omega_{b0}=0.038^{+0.002}_{-0.008}$. The value of $H_{0}$  obtained is close to the one reported by  Riess et al \cite {H0}, $H_{0} = 74.2 \pm 3.6 {\rm km s^{-1} Mpc^{-1}}$ at $68 \%$ C.L.,  being the measurement of $H_{0}$ obtained from the magnitude–-redshift
relation of 240 low-$z$ Type Ia supernovae at $ z < 0.1$; the absolute magnitudes of supernovae are calibrated using new observations 
from Hubble Space Telescope (HST) of 240 Cepheid variables in six local Type Ia supernovae host galaxies and the maser galaxy NGC 4258. Further, a 7-year WMAP analysis prefers, but does not directly measure, $H_{0} = 71.0 \pm 2.5 {\rm km s^{-1} Mpc^{-1}}$ \cite{WMAP7}. The value of the Hubble constant  was evaluated over a range of redshifts $0.03 < z < 0.5$, assuming a cosmological model with $\Omega_{c0} = 0.27$ and $\Omega_{x0} = 0.73$. The value determined for the Hubble constant is  $H_{0} =75.9 \pm 3.8 {\rm km s^{-1} Mpc^{-1}}$ \cite{SNH0}. Besides, the value of $\Om_{c0}$ is consistent with $\Om_{c0}h^{2}=0.11$ and $h=0.7$ \cite{WMAP7}.  We will use these mean values to analyze the main traits of the model with the third component, in particular the issue of early dark energy will be addressed with some detail as a way of further constrain our model with the physics  behind  the primordial eras such as recombination or  big bang nucleosynthesis.
\end{widetext}

\begin{center}
\begin{table}
\begin{minipage}{1\linewidth}
\scalebox{0.84}{
\centering %
\begin{tabular}{lll}
\hline
\multicolumn{2}{c}{2D Confidence level} \\
\hline
Priors & Best fits\\
\hline
{$(\Omega_{m0},\Omega_{x0}, \beta)=(0.21, 0.75, 0.005)$} & $( \al, h)=(1.089^{+0.089}_{-0.134}, 0.835^{+0.005}_{-0.002})$\\
{$(\Omega_{m0},\Omega_{x0}, \al)=(0.21, 0.75, 1.0887)$}& $( \beta, h)=(0.005^{+0.065}_{-0.111}, 0.835^{+0.005}_{-0.002})$\\
{$(\Omega_{x0}, \al, \beta)=( 0.75, 1.0887, 0.005)$}& $(\Omega_{m0}, h)=(0.21^{+0.40}_{-0.35}, 0.835^{+0.003}_{-0.002})$\\
{$(\Omega_{m0}, \al, \beta)=( 0.21, 1.0887, 0.005)$}& $(\Omega_{x0}, h)=(0.75^{+0.03}_{-0.02}, 0.835^{+0.004}_{-0.002})$\\
{$(\Omega_{m0}, \al, h)=( 0.21, 1.0887, 0.835)$}& $(\Omega_{x0}, \beta)=(0.75^{+0.08}_{-0.06}, 0.005^{+0.188}_{-0.181})$\\
{$(\Omega_{x0}, \al, h)=( 0.75, 1.0887, 0.835)$}& $(\Omega_{m0}, \beta)=(0.21^{+0.52}_{-0.52}, 0.005^{+0.076}_{-0.077})$\\
{$(\Omega_{m0}, \beta, h)=( 0.21, 0.005, 0.835)$}& $(\Omega_{x0}, \al)=(0.75^{+0.06}_{-0.32},  1.089^{+0.551}_{-0.589})$\\
{$(\al, \beta, h)=( 1.0887, 0.005, 0.835)$}& $(\Omega_{x0}, \Omega_{m0})=(0.75^{+0.04}_{-0.04},  0.21^{+0.76}_{-0.86})$\\
{$(\Omega_{m0},\Omega_{x0}, h)=(0.21, 0.75, 0.835)$} & $( \al, \beta)=(1.089^{+0.186}_{-1.549}, 0.005^{+0.440}_{-1.139})$\\
\hline
\end{tabular}}
\caption{\scriptsize{ Observational bounds for the 2D C.L. obtained in  Fig. (\ref{Fig1b}) by varying two cosmological parameters. It is reported  the best fit values of the cosmological parameters with theirs corresponding  marginal  $1\sigma$  error-bars.}}
\label{snet}
\end{minipage}
\end{table}
\end{center}

\begin{figure}[hbt!]
\begin{minipage}{1\linewidth}
\resizebox{1.6in}{!}{\includegraphics{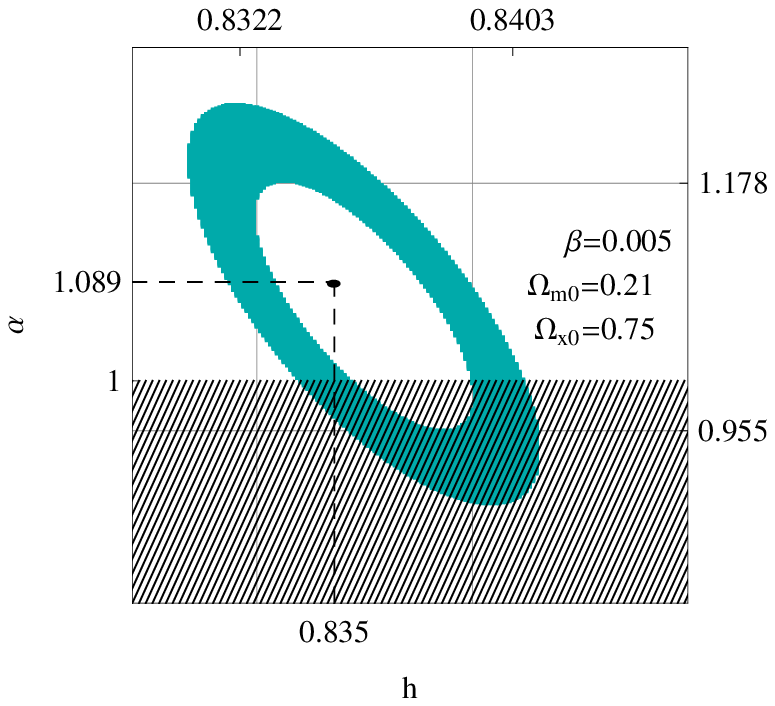}} 
\resizebox{1.6in}{!}{\includegraphics{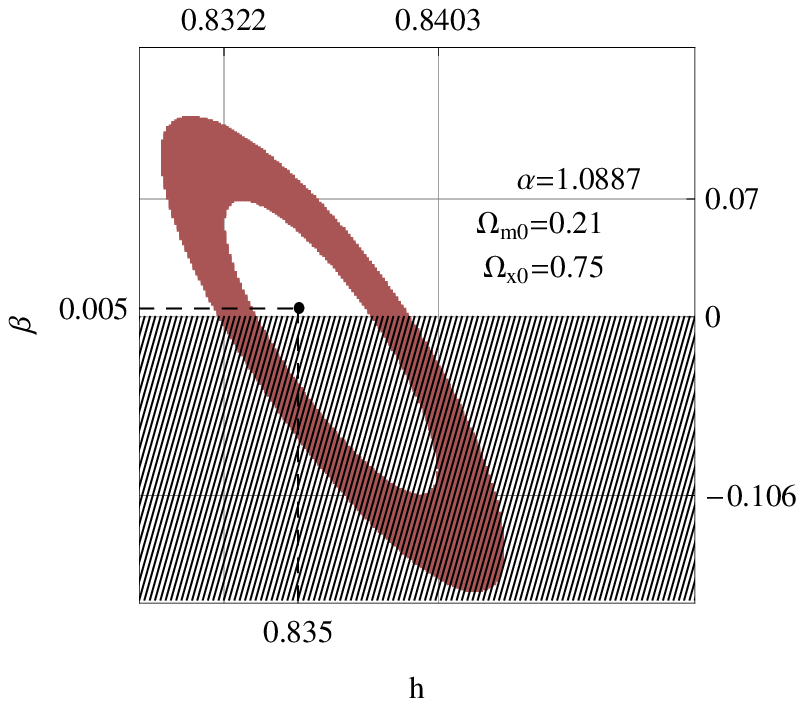}}\hskip0.05cm 
\resizebox{1.6in}{!}{\includegraphics{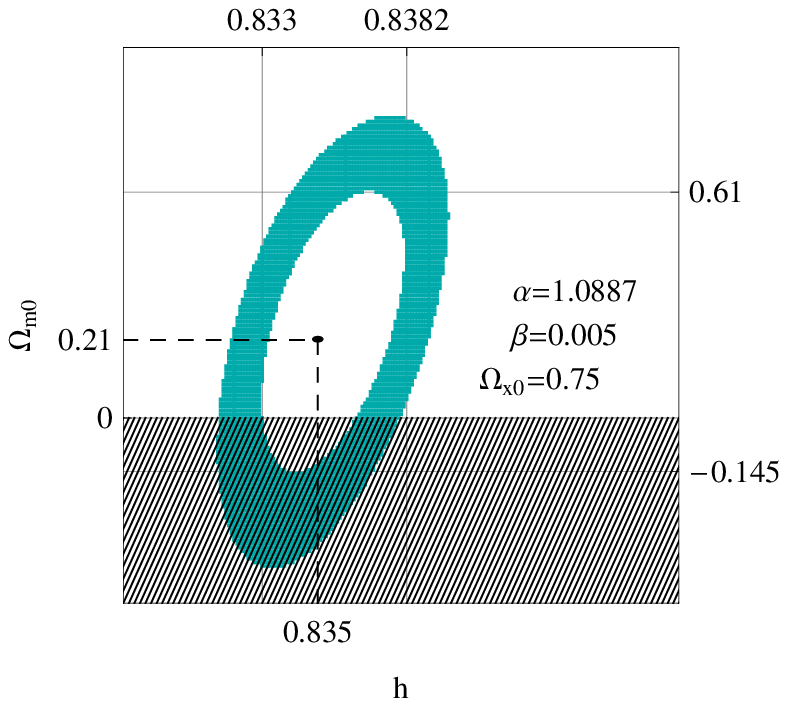}}\hskip0.05cm 
\resizebox{1.6in}{!}{\includegraphics{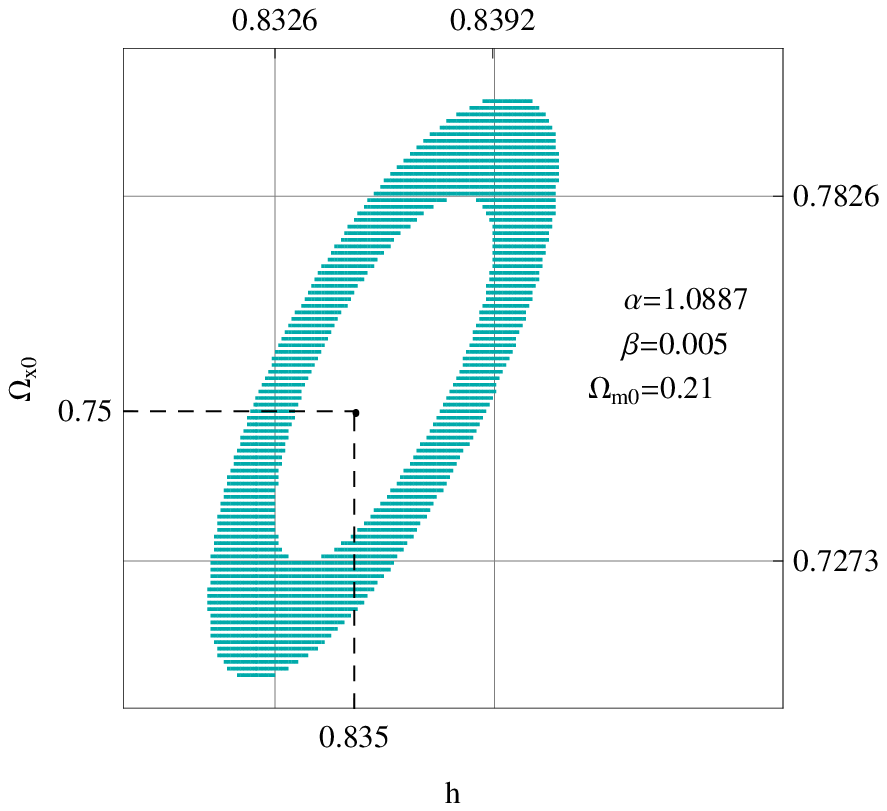}}
\resizebox{1.6in}{!}{\includegraphics{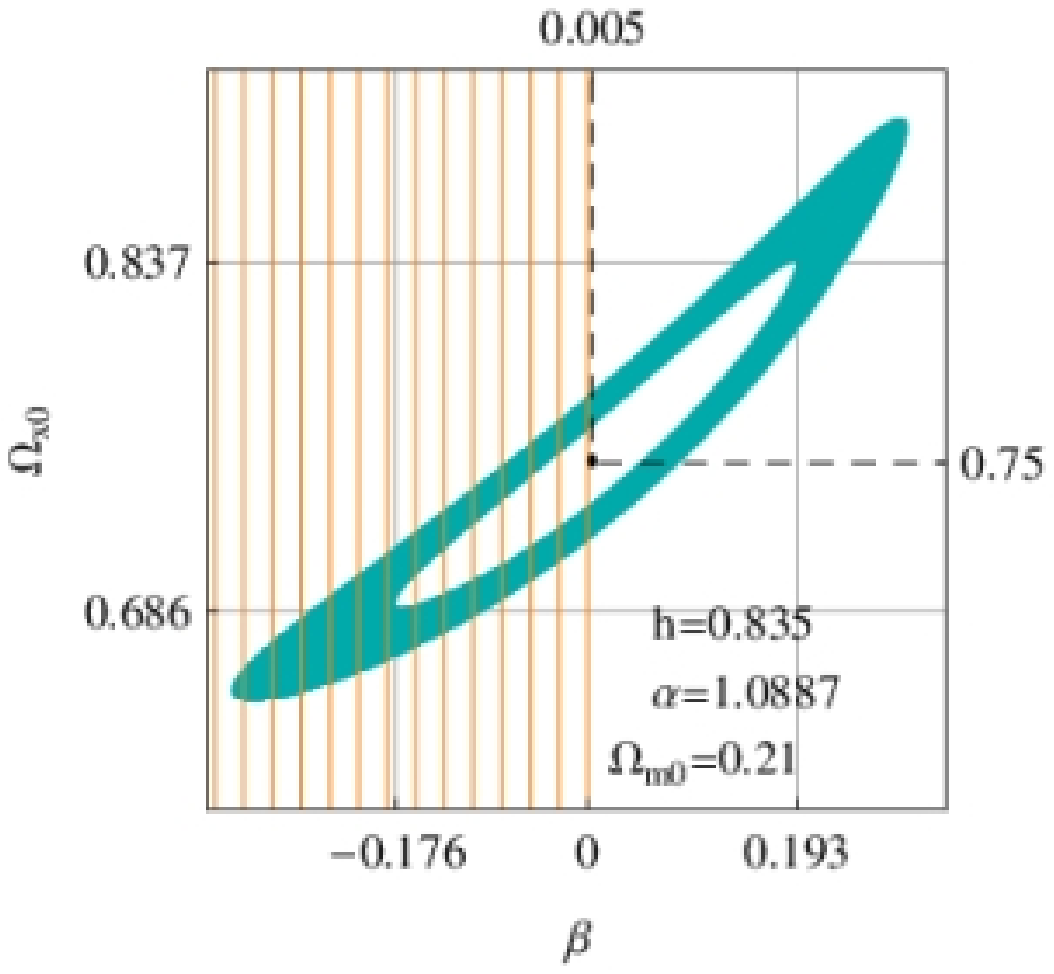}}
\resizebox{1.6in}{!}{\includegraphics{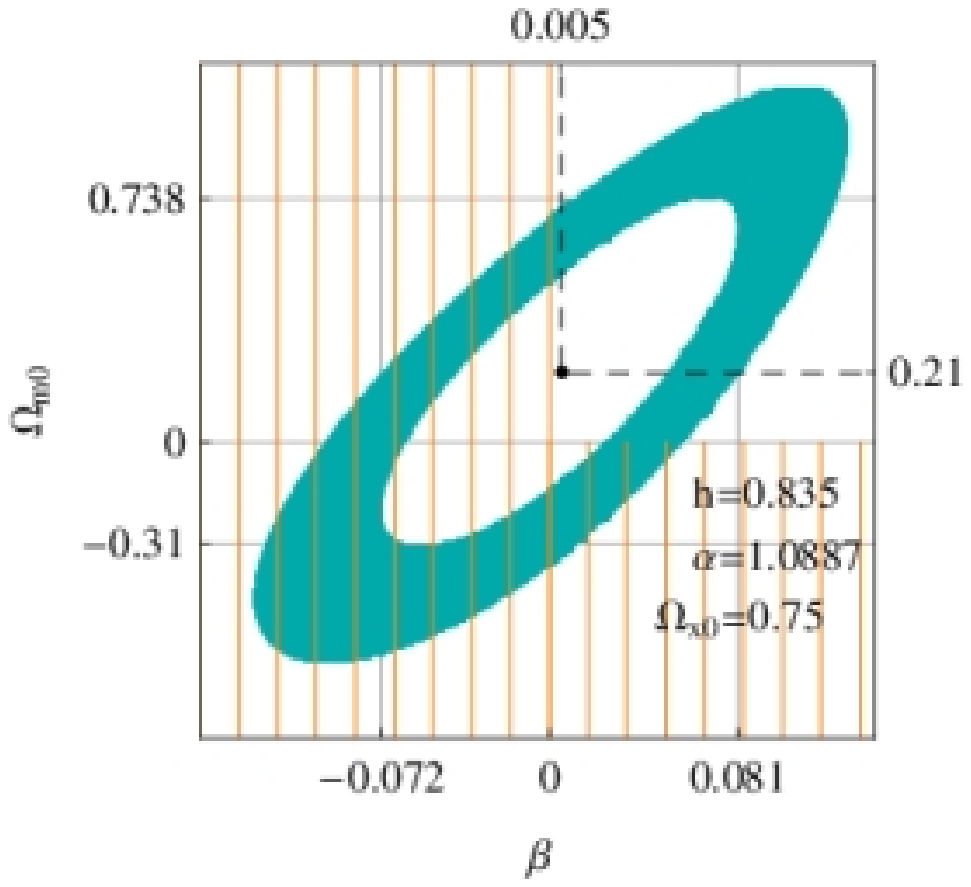}}
\resizebox{1.6in}{!}{\includegraphics{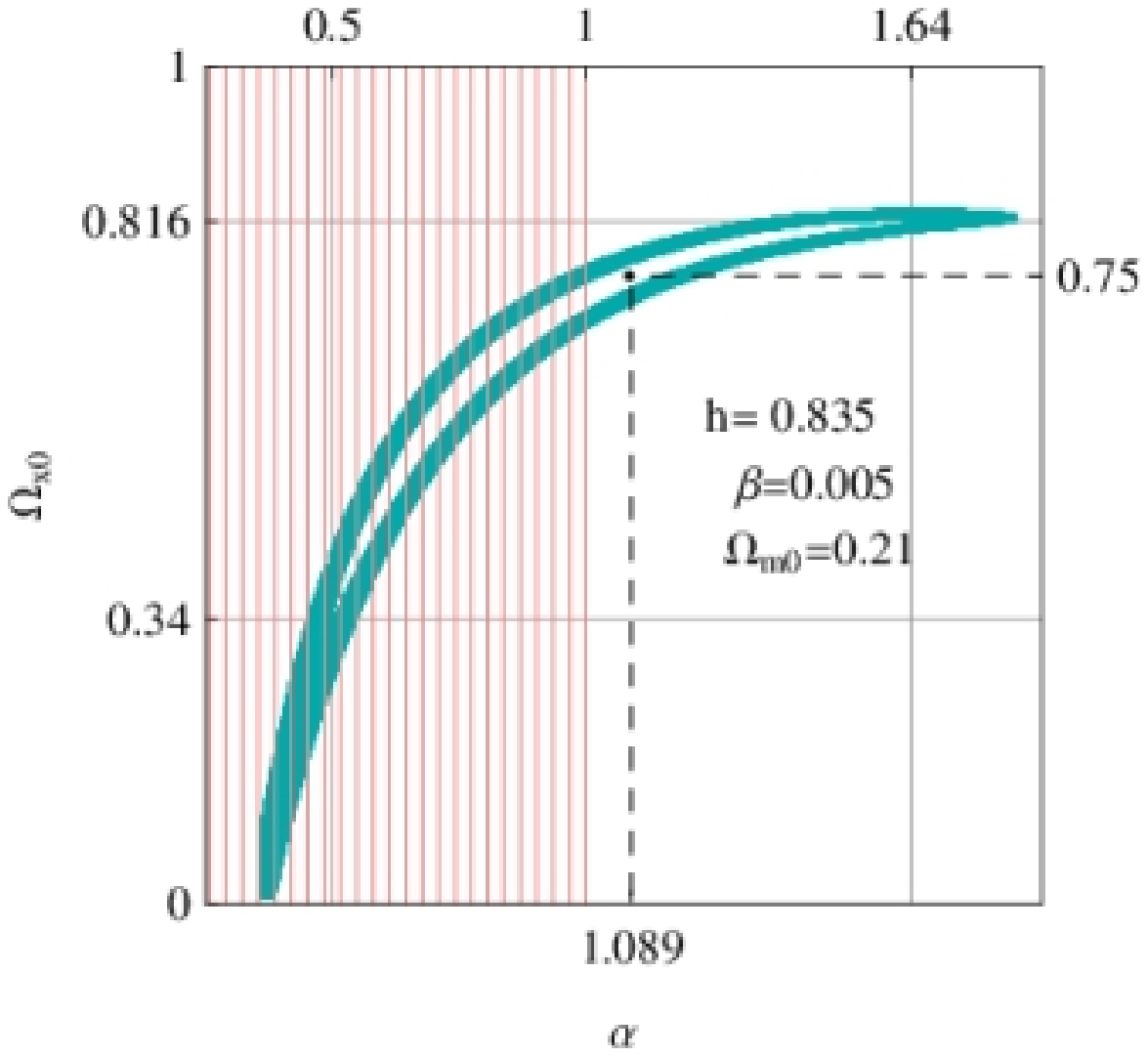}}
\resizebox{1.6in}{!}{\includegraphics{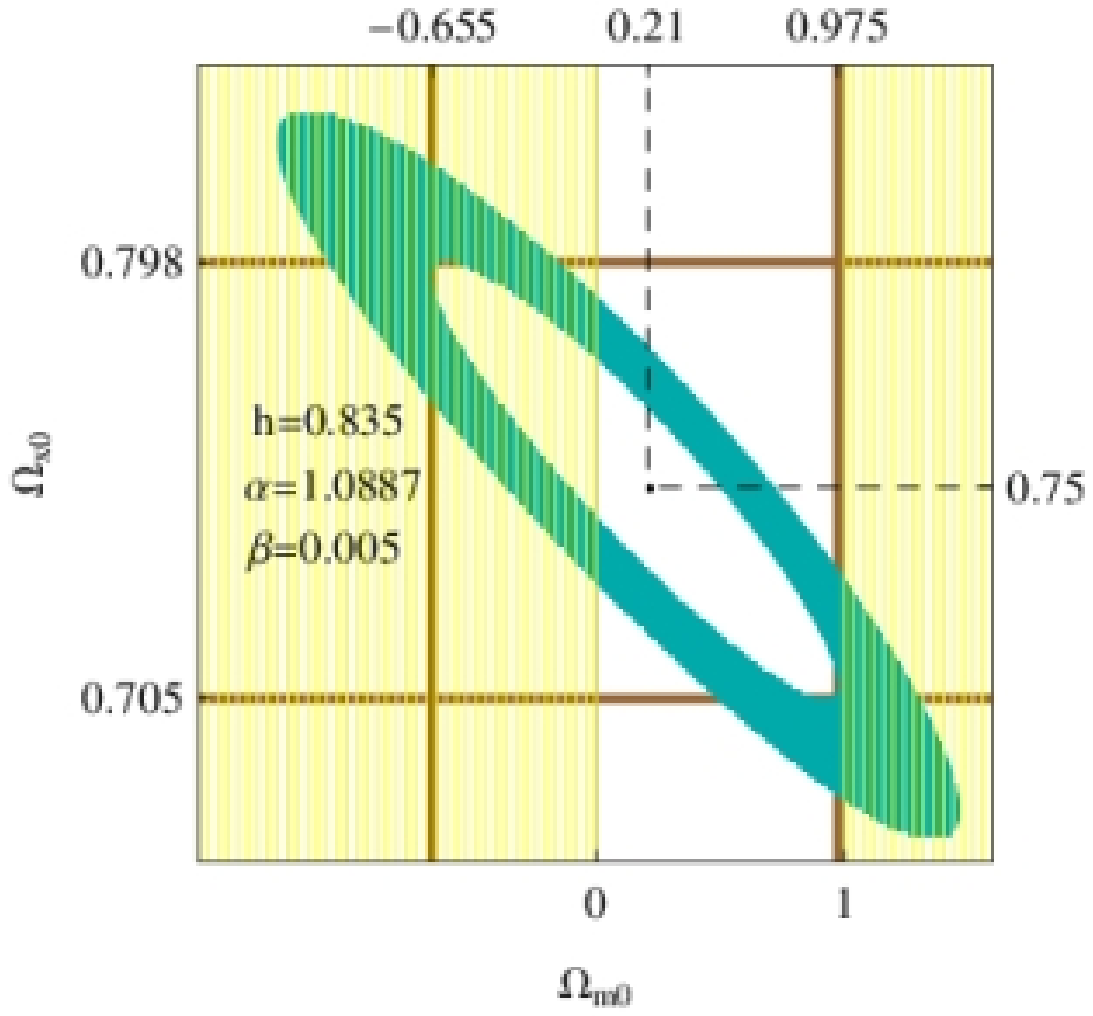}}
\resizebox{1.6in}{!}{\includegraphics{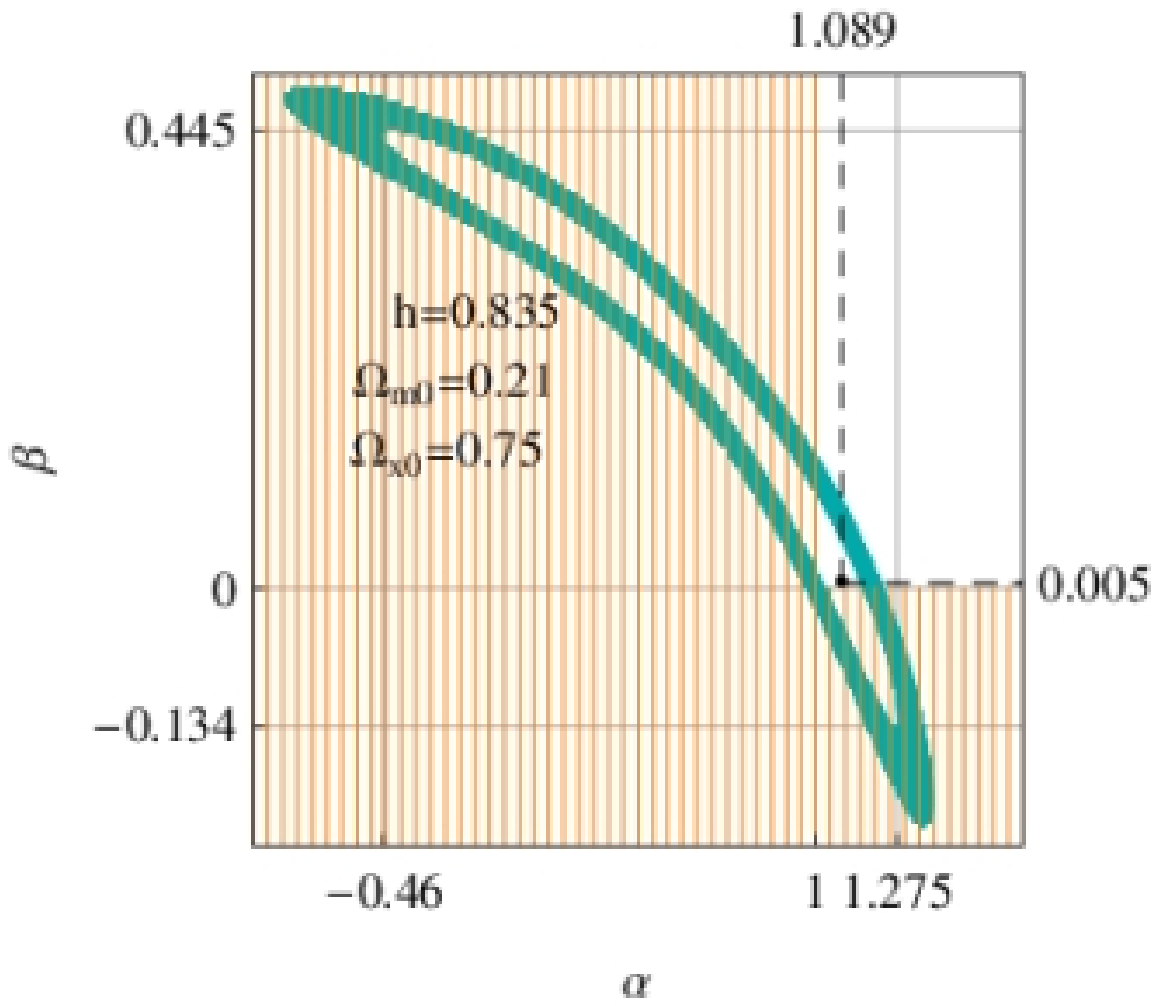}}
\caption{\scriptsize{Two-dimensional C.L. associated with $1\sigma$ and $2\sigma$ for different $\theta$ planes obtained to the Union 2 compilation of SNe Ia.}}
\label{Fig1b}
\end{minipage}
\end{figure}

As is well known, distance indicators can be used for confronting distance measurements to the corresponding
model predictions. One of the most useful ones are those objects of known
intrinsic luminosity such as  standard candles, so that the corresponding comoving
distance can be determined. That way, it is possible to reconstruct the Hubble expansion rate by searching this
sort of object at different redshifts. The most important
class of such indicators is type Ia supernovae.  Then, we would like to   compare the Hubble data  with the
Union2 compilation of 557 SNe Ia \cite{amanu}. In order to do that, we note that the apparent magnitude of a supernova placed at a given
redshift z is related to the expansion history of the Universe through the distance modulus
\be
\n{mu}
\mu\equiv m -M= 5\log \frac{d_{L}(z)}{h}+\mu_{0},
\ee
where $m$ and $M$ are the apparent and absolute magnitudes, respectively, $\mu_{0}=42.38$, $h=H_{0}/100\rm{km^{-1} s^{-1}}$, and $d_{L}(z)=H_{0}(1+z)r(z)$, being $r(z)$ the comoving distance, given for a FRW metric by 
\be
\n{r}
r(z)=\int^{z}_{0}{\frac{dz'}{H(z')}}.
\ee
To confront the model with supernovae data set we construct the corresponding $\chi^2$ estimator
\be
\n{c1p}
\chi^2(\theta) =\sum_{k=1}^{N}\frac{[\mu(\theta,z_k) - \mu(z_k)]^2}{\sigma(z_k)^2},
\ee
where $N=557$ and the cosmological parameters are  $\theta= (H_0, \al, \bb,\Om_{c}, \Om_{x}, \Om_{b})$. Using the Union2 data set, we will obtain nine two-dimensional confidence contours associated to $1\sigma$  and $2\sigma$ error  (see Fig. (\ref{Fig1b})). Thus,  we obtain the best fit values for nine cases and calculate the corresponding marginal $1\sigma$ error bars \cite{Bayes} as it can be  seen in Table (\ref{snet}).  The dashed zones are  excluded from the analysis due to different reasons such as it can be that the range of the parameters $\al$ and $\beta$ lead to a phantom scenario, the parameter densities $\Omega_{i0}$ become negative or take values that are not consistent with the literature.  More precisely, we found that the values of $\al$ vary over the interval  $[1.08, 1.64]$ within $1\sigma$ zone whereas $\beta \in [0.005; 0.445]$. At $1\sigma$ C.L. the dark energy density parameter at $z=0$ goes from 0.34 to 0.816, $\Omega_{m0} \in [0.21, 0.975]$, and $h \in [0.832; 0.8350]$ (see Table(\ref{snet})). The difference between the forecast made with the  Hubble data and Union2 set  is most sharpest in the case of $\beta$ parameter, it  exhibits a disagree of $0.9\%$. For $\rm h$ and $\Omega_{m0}$ the discrepancy between both set does not reach $0.14\%$ whereas the values of  $\rm \al $ and $\Omega_{x0}$ obtained with Hubble data disagree with the ones of Union2 by $0.03 \%$.
In order to corroborate  our previous analysis, we also  performed a  global statistical analysis with  the Union2 data set by taking into account a global minimization of the five parameters. The latter procedure leads to the best fit values $(h, \al, \bb,\Om_{c}, \Om_{x})=(0.83,1.08,0.005,0.21,0.75)$ along with $\chi^2_{\rm d.o.f}= 0.98<1$, indicating that our estimations of the cosmological parameters are  trustworthy.  The statistical estimations performed  with the Union2 data set are also consistent with the ones obtained from the Hubble data set.

The density parameters $\Omega_{c}$, $\Omega_{x}$, $\Omega_{b}$, and the ratio $r=\Omega_{c}/\Omega_{x}$ in term of the redshift $z$ are given by
\be
\n{28a}
\Om_c =\frac{1}{\Delta}\left(\frac{(1-\bb){\cal A}+\bb(\al-1){\cal B}(1+z)^{3(\al\bb-1)}}{{\cal A}+ {\cal B}(1+z)^{3(\al\bb-1)}+ \Omega_{b0}(1+z)^{3(\al-1)}}\right),  
\ee
\be
\n{29}
\Om_b =\frac{\Omega_{b0}(1+z)^{3(\al - 1)}}{ {\cal A}+ {\cal B}(1+z)^{3(\al\bb-1)}+ \Omega_{b0}(1+z)^{3(\al-1)}}, 
\ee
\be
\n{28b}
\Om_x =\frac{1}{\Delta}
\left(\frac{(\al-1){\cal A}+\al(1-\bb){\cal B}(1+z)^{3(\al\bb-1)}}{{\cal A}+{\cal B}(1+z)^{3(\al\bb-1)}+ \Omega_{b0}(1+z)^{3(\al-1)}}\right),   
\ee
\be
\n{28c}
r=\frac{(1-\bb){\cal A}+\bb(\al-1){\cal B}(1+z)^{3(\al\bb-1)}}{ {\cal A}+ {\cal B}(1+z)^{3(\al\bb-1)}+ \Omega_{b0}(1+z)^{3(\al-1)}}
\ee
\vskip0.3cm
 
\begin{figure}[hbt]
\begin{minipage}{1\linewidth}
\resizebox{3.0in}{!}{\includegraphics{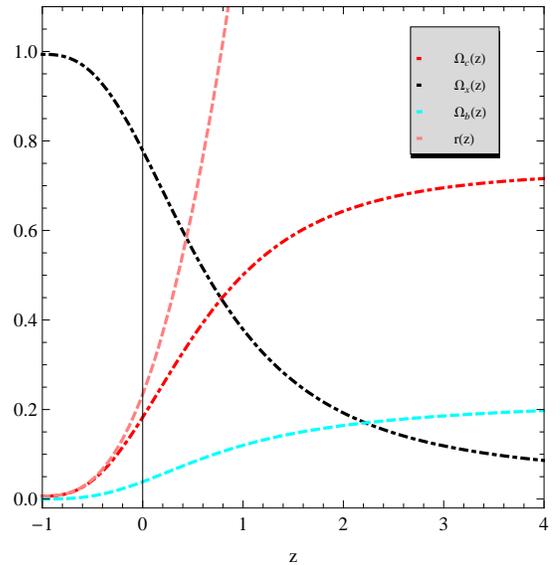}}
\caption{\scriptsize{The density parameters  $\Omega_c$, $\Omega_x$, and $\Omega_b$ as a function of the redshift are depicted. Also the ratio $\Omega_c/\Omega_x$  is shown.}} 
\label{Fig:Figura10}
\end{minipage}
\end{figure}

The aforesaid model (\ref{22}) exhibits  dark matter and dark energy components  with energy densities of similar order of magnitude at redshifts $z \leq 2$ including the present-day scenario (see Fig.\ref{Fig:Figura10}).
Now, we are interested in studying the behavior of kinematic quantities such as deceleration parameter and the equations of state.
Indeed,  we readily get to obtain $\om(z)$, $\om_{c}=0$, $\om_{x}(z)$, $\om_{b}=\al-1$, and $q(z) = [1+3\om(z)]/2$ in term of the redshift  
\be
\n{30a}
\om =-1+\left[\frac{ {\cal A}+\al\bb {\cal B}(1+z)^{3(\al\bb-1)}+\al\Om_{b0}(1+z)^{3(\al-1)}}{ {\cal A}+ {\cal B}(1+z)^{3(\al\bb-1)}+ \Omega_{b0}(1+z)^{3(\al-1)}}\right],  
\ee
\be
\n{30b}
\om_x =\frac{1}{\Om_x}\left[\om - \frac{(\al-1)\Om_{b0}(1+z)^3}{(H/H_0)^2}\right].   
\ee
It seems like the  model experiences a dust-like behavior around $z = 7.3$ and the transition towards the accelerated regime takes place at  $z=0.87$ so it is retarded when the nonintercating baryonic matter is added. Besides, the actual values of all equations of states and the deceleration parameter are $\om_{0}=-0.63$, $\om_{c0}=0$, $\om_{x0}=-0.82$, $\om_{b0}=0.06$, and $q_0=-0.45$, respectively. These values are   consistent with the high-$z$ supernova data  which provide the most stringent limit on $\omega_{x0}$. Using WMAP+BAO+SN, it was obtained $\omega_{x0} = -0.980 \pm
0.053$ ($68\%$ C.L.). The error does not include systematic errors in supernovae, which are comparable to the statistical error, thus, the error in $\omega_{x0}$ from WMAP+BAO+SN is about a half of that from WMAP+BAO+$H_{0}$ \cite{WMAP7}.  The model does not cross phantom divide line at any stage of its evolution, moreover in the remote future decreases monotonically reaching a value of $\om_{x0} \simeq - 0.82$, in the same way as happens for a quintessence dark energy model, so it does not exhibit a quintom behavior \cite{NQ}. 
\begin{figure}[hbt]
\begin{minipage}{1\linewidth}
\resizebox{4.5in}{!}{\includegraphics{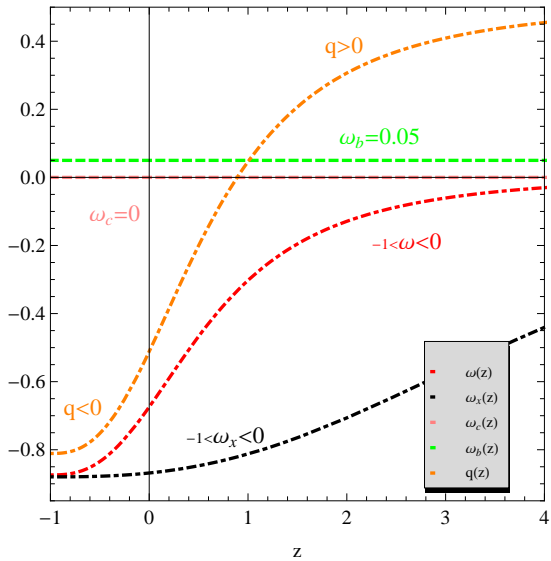}}
\caption{\scriptsize{ It shows  $\om(z)$, $\om_{c}(z)$, $\om_{x}(z)$, $\om_{b}(z)$, and $q(z)$ in term of the redshift $z$.}} 
\label{Fig:Figura12}
\end{minipage}
\end{figure} 

\begin{figure}[h!]
\begin{center}
\includegraphics[height=9.5cm,width=8.4cm]{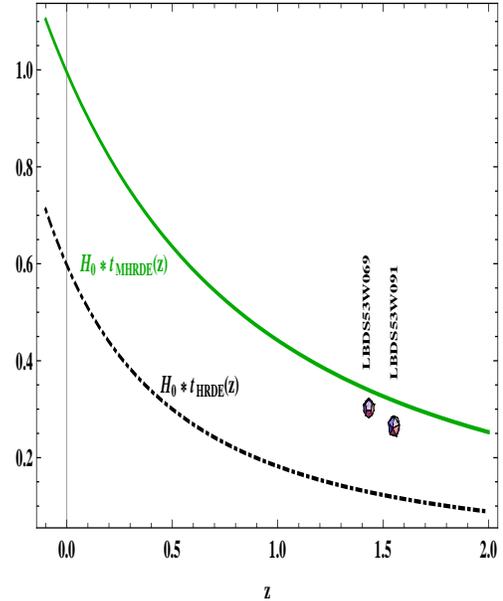}
\caption{It shows the cosmic age-redshift relation for the MHRDE with and without  the noninteraction baryonic matter ( dashed green lines),  HRDE  (dashed black curve). }
\label{Fig:Figura13}
\end{center}
\end{figure}

The cosmic age-redshift relation for our model reads
\be
\n{31}
H_0 t(z)= \int_z^{\infty} \frac{dx}{(1+x)(H(z)/H_0)},
\ee
\no where the time origin is set at $z = \infty$ and the time is measured in units of $H_0$. For the best fit parameters, we found that the age of universe is $13.31 {~\rm Gyr}$  without the third fluid whereas its inclusion leads to  $13.17 {~\rm Gyr}$. Both values are very close to the one reported by  WMAP-7year project, thus it found a $13.75 \pm 0.13 {~\rm Gyr}$ with $WMAP$ only and $13.75\pm 0.11{~\rm Gyr}$ with $WMAP +BAO+H_{0}$ \cite{Jarosik:2010iu}. Because the cosmological constraints with the Hubble data only cover redshifts over the range $0 \leq z<2$, the comparison with cosmic milestones will be trustworthy in this range only, for the latter reason we consider  two old stellar sources such as the $4 {~\rm Gyr}$ old galaxy LBDS 53W069 at redshift $z = 1.43$ \cite{Dunlop:1997} and  the $3.5 {~\rm Gyr}$ old galaxy LBDS 53W091 at redshift $z = 1.55$ \cite{Dunlop:1996mp} [see  Fig.\ref{Fig:Figura13}]. We depict the age-redshift relations  at the best-fit value corresponding to the two models mentioned before (see  Fig.\ref{Fig:Figura13}). We find that the  Ricci-like holographic dark energy model cannot be accomodated well under the age-redshift curves  exhibiting a cosmic-age problem at low redshift, namely,   the universe cannot be younger than its constituents. On the other hand,  the MHRDE  seems to be free from the cosmic-age problem at low redshift.

\begin{figure}[h!]
\begin{center}
\includegraphics[height=9cm,width=9.4cm]{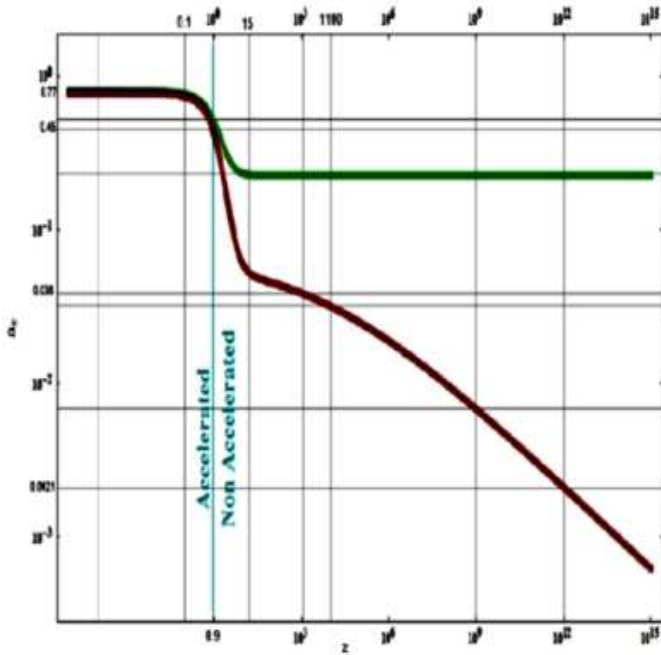}
\caption{The cosmic behavior of  the dark energy density  $\Omega_x$  in term of the redshift over the interval $[0, 10^{15}]$ for the MHRDE model. The green solid line shows the case where the universe is filled with dark matter and dark energy, whereas  the red solid line  indicates a universe filled with three components, two of them  encoded in the interacting dark sector, and the third one  is a baryonic non interacting fluid. }
\label{Fig:Figura14}
\end{center}
\end{figure}

Now, we would like to attend an  appealing discussion concerning the behavior of dark energy at early times  within the framework of interacting dark sector for the two cases; one case corresponds to the interacting dark sector only, whereas the second one is the model with the three components.  In doing so, we exhibit the cosmological evolution of the density parameters $\Omega_x (z)$  for the first model when the universe is filled  with dark matter and dark energy only, whilst the second model refers to a Universe  filled with  an interacting dark sector plus a non interacting baryonic fluid [see Eq. (\ref{28b})]. The dark energy is depicted over the range $z \in [0, 10^{15}]$ in order to have in mind both the bounds coming from a recombination era as well as  those produced by BBN data. The idea is to compare the behavior of dark energy at early times for the two  models mentioned in this article, and thus, we also will contrast our finding with the ones reported in the literature. Although both models  have the same kind of interaction it turned out that the behavior of  their density  parameters  are very different at early times (see Fig.\ref{Fig:Figura14}). In broad terms, both models exhibit a stable behavior about the current value $0.77$ at low redshifts ( $z < 0.1$) and then they begin to separate abruptly  within the interval $ 0.1 < z < 15 $, showing a descending slope.   If  the non-interacting fluid is excluded, the density parameter $\Omega_{x}$ exhibits another plateau very similar to one of the late times but now the amount of dark energy  is fixed around $0.23$, so it would not strictly satisfy the recombination bound, nevertheless at $z\simeq 10^{8}$ corresponding to a BBN era it would be close to meet the condition $\Omega_{x}(z \simeq 10^{10})<0.21$.  On the other hand, the inclusion of baryonic non interacting fluid makes possible that the amount of dark energy  continues descending rapidly, exhibiting values which are perfectly in agreement with the stringent constrains provided by the EDE at recombination era.  For instance $\Omega_x (z=10^3) =0.038$ and $\Omega_x (z\simeq 1100) =0.032$, these values  indicate that the second model is consistent with the forecast of  Planck  and CMBPol experiments \cite{EDE2}, meeting below the  upper bound provided by  the constraints on the variation in the fine structure constant \cite{EDE4} and showing a slightly discrepancy of one order of magintude  with the  bound reported in \cite{EDE3} when the CMB data alone 
is used. As we have already mentioned, the presence of dark energy  at BBN era should not disturb the observed Helium abundance in the universe which is regarded as one of the major evidence in supporting the big bang theory. In relation with that, we  found that the amount of dark energy  is $\Omega_{x}(z\simeq 10^{10})<0.21$ at BBN \cite{Cyburt}. In our model, we obtained  $\Omega_x (z=10^9) =0.0068$, and $\Omega_x (z=10^{12}) =0.0021$ so the fraction of dark energy at early times clearly fulfills the aforesaid constraint  (see Fig.\ref{Fig:Figura14}). It should be also  stressed that the most importants changes are produced near the transition era when  the universe entered in the accelerated regime (Fig.\ref{Fig:Figura14}). In addition, we find that within the framework of holographic Ricci dark energy model the bound at recombination era is not satisfied with the baryonic component or without it. In a future research, we will explore this kind of interaction  by taking a radiation or baryonic term coupled to the dark sector;  we will examine the changes introduced in the behavior of dark energy at early times \cite{tfmi}.

Finally, notice that the value of the cosmological parameters used here are not arbitrary  because these parameters turned to be consistent with three important data set: $\rm i-$ the present-day scenario constrained with the Hubble data and SNe Ia data set, $\rm ii-$ the recombination bounds for EDE, and  $\rm iii-$ the BBN data. Another useful observational constraints can be found in the last scattering surface (LSS), namely, during the galaxy formation era ($1<z<3$) dark energy density have to be subdominant to matter density  so it should satisfy that $0<\Omega_{x}<0.5$ as happens in our case. 

\section{Conclusions}

We have  examined a FRW universe  filled with dark matter, a MHRDE with a cutoff  given by  $L^{-2} = (\dot H + \frac{3\al}{2} H^2)/(\al-\bb)$   that includes Ricci ansatz for $\al=4/3$, and a baryonic non interacting component decoupled from the dynamics of dark sector.  We have studied the case in which the interaction in the dark sector is proportional to its energy density derivative, namely, $Q\propto\ro'$ and found that the total pressure of both dark components becomes strongly negative, violating the strong energy condition in the present epoch.

We have  performed a $\chi^2$-analysis using the Hubble data and built the corresponding $1\sigma$ and $2\sigma$ C.L. (see Fig.\ref{Fig: Figura9}) for each pair of parameters (see Table \ref{Tabla3}). The mean values of $\theta_{c}=(H_{0}, \al, \bb, \Omega_{x0}, \Omega_{c0}, \Omega_{b0})$ are $H_0=73.89^{+0.44}_{-0.94}{\rm km~s^{-1}\,Mpc^{-1}}$, $\alpha=1.055^{+0.015}_{-0.038}$, $\beta=0.126^{+0.044}_{-0.096}$, $\Omega_{c0}=0.18^{+0.01}_{-0.01}$, $\Omega_{x0}=0.768^{+0.002}_{-0.008}$ and $\Omega_{b0}=0.038^{+0.002}_{-0.008}$ along with a $\chi^{2}_{d.o.f}<1$. Taking into account these $\theta_{c}$ into the density parameters of the dark components, we have found that they have similar behavior for redshifts $z \leq 2$ including the present-day scenario (see Fig.\ref{Fig:Figura10}). Regarding the statistical estimations made with the Union 2 compilation of SNe Ia, we have found that $\al$ varies over the interval  $[1.08, 1.64]$ within $1\sigma$ region whereas $\beta \in [0.005; 0.445]$. At $1\sigma$ C.L. the dark energy density parameter at $z=0$ goes from 0.34 to 0.816, $\Omega_{m0} \in [0.21, 0.975]$, and $h \in [0.832; 0.8350]$ (see   Fig. (\ref{Fig1b}) and Table(\ref{snet})).  We have also performed a  global statistical analysis with  the Union2 data set by taking into account a global minimization of the five parameters. The latter procedure leads to the best fit values $(h, \al, \bb,\Om_{c}, \Om_{x})=(0.83,1.08,0.005,0.21,0.75)$ along with $\chi^2_{\rm d.o.f}= 0.98<1$, indicating that the estimation made with the Union2 data  is  trustworthy and  consistent with the one obtained from the Hubble data set.

The kinematic analysis based on the behavior of deceleration parameter indicates that at $z_{t}=0.87$ the universe begins to accelerate, being $q_0=-0.45$ its current value within $1\sigma$ C.L..  Concerning the equations of state, we have found that $-1<\omega_{x}(z), \omega(z)<0$, $\omega_{c}=0$, and $\omega_{b}=\al-1>0$  along with theirs current values $\om_{0}=-0.63$, $\om_{c0}=0$, $\om_{x0}=-0.82$, and $\om_{b0}=0.05$ (cf. Fig. \ref{Fig:Figura12}).  In addition, we have obtained that the age of the universe is $13.17 {~\rm Gyr}$  very close to the one reported by  WMAP-7year project or WMAP +BAO+$H_{0}$ data \cite{Jarosik:2010iu}. So that the MHRDE  is free from the cosmic-age problem at low redshift($0 \leq z<2$) in contrast to the Ricci-like HDE, giving rise age-redshift curves below  two old stellar sources such as the $4 {~\rm Gyr}$ old galaxy LBDS 53W069 at redshift $z = 1.43$ \cite{Dunlop:1997} and  the $3.5 {~\rm Gyr}$ old galaxy LBDS 53W091 at redshift $z = 1.55$ (cf. Fig. \ref{Fig:Figura13}).

We have studied the issue of dark energy at early times by taking into account the stringent bounds reported at recombination era and/or  at BBN and shown that the inclusion of a non interacting component makes possible that the amount of dark energy at  $z_{t} \sim {\cal O}(1)$ begins to decrease sharply, giving  $\Omega_x (z=10^3) =0.038$ and $\Omega_x (z\simeq 1100) =0.032$  at recombination era. These bounds indicate a good agreement with the forecast of  Planck  and CMBPol experiments \cite{EDE2} as well as  with the upper bound provided by  the constraints on the variation in the fine structure constant \cite{EDE4}. As we have already mentioned, the presence of dark energy  at BBN era should not disturb the observed Helium abundance in the universe which is regarded as one of the major evidence in supporting the big bang theory. We have obtained the dark energy density parameters $\Omega_x (z=10^9) =0.0068$ and $\Omega_x (z=10^{12}) =0.0021$ so the fraction of EDE  fulfills the bound $\Omega_{x}(z\simeq 10^{10})<0.21$ at BBN \cite{Cyburt} (see Fig.\ref{Fig:Figura14}). Finally, we would like to stress that  the values of the cosmological parameters obtained here are consistent with four important data set: $\rm i-$ the present-day scenario obtained with the Hubble data and SNe Ia data set, $\rm ii-$ the recombination bounds for EDE, $\rm iii-$ the BBN data, and $\rm iv-$ LSS.

\section{acknowledgments}
We would like to thank the referee for making useful suggestions which  helped improve the article.
L.P.C thanks  the University of Buenos Aires  for their support under Project No. 20020100100147 and the Consejo Nacional de Investigaciones Cient\'{\i}ficas y T\' ecnicas (CONICET) through the research Project PIP 114-200801-00328. M.G.R   is partially supported by  Postdoctoral Fellowship Programme of  CONICET.

\end{document}